**Disentangling magnetic and grain contrast in polycrystalline FeGe thin films using four-dimensional Lorentz scanning transmission electron microscopy**

Kayla X. Nguyen[1,2*], Xiyue S. Zhang[2*], Emrah Turgut[2,3], Michael C. Cao[2], Jack Glaser[2], Zhen Chen[2], Matthew J. Stolt[4], Celesta S. Chang[5], Yu-Tsun Shao[2], Song Jin[4], Gregory D. Fuchs[2,6], David A. Muller[2,6]

1. Department of Chemistry and Chemical Biology, Cornell University, Ithaca, NY 14853, USA
2. School of Applied and Engineering Physics, Cornell University, Ithaca, NY 14853, USA
3. Department of Physics, Oklahoma State University, Stillwater, OK 74078, USA
4. Department of Chemistry, University of Wisconsin–Madison, Madison, Wisconsin 53706, USA
5. Department of Physics, Cornell University, Ithaca, NY 14853, USA
6. Kavli Institute at Cornell for Nanoscale Science, Ithaca, NY 14853, USA

*These Authors Contributed Equally

Abstract

The study of nanoscale chiral magnetic order in polycrystalline materials with a strong Dzyaloshinkii-Moriya interaction (DMI) is interesting for the observation of magnetic phenomena at grain boundaries and interfaces. This is especially true for polycrystalline materials, which can be grown using scalable techniques and whose scalability is promising for future device applications. One such material is sputter-deposited B20 FeGe on Si, which has been actively investigated as the basis for low-power, high-density magnetic memory technology in a scalable material platform. Although conventional Lorentz electron microscopy provides the requisite spatial resolution to probe chiral magnetic textures in single-crystal FeGe, probing the magnetism of sputtered B20 FeGe is more challenging because the sub-micron crystal grains add confounding contrast. This is a more general problem for polycrystalline magnetic devices where scattering from grain boundaries tends to hide comparably weaker signals from magnetism. We address the challenge of disentangling magnetic and grain contrast by applying 4-dimensional Lorentz scanning transmission electron microscopy using an electron microscope pixel array detector. Supported by analytical and numerical models, we find that the most important parameter for imaging magnetic materials with polycrystalline grains is the ability for the detector to sustain large electron doses, where having a high-dynamic range detector becomes extremely important. Despite the small grain size in sputtered B20 FeGe on Si, using this approach we are still able to observe helicity switching of skyrmions and magnetic helices across two adjacent grains as they thread through neighboring grains. We reproduce this effect using micromagnetic simulations by assuming that the grains have distinct orientation and magnetic chirality and find that magnetic helicity couples to crystal chirality. Our methodology for imaging magnetic textures is applicable to other thin-film magnets used for spintronics and memory applications, where an understanding of how magnetic order is accommodated in polycrystalline materials is important.

## I. Introduction

Power-efficient memory devices based on magnetic skymions in sputtered, polycrystalline thin films are increasingly promising [1-5]. Although the crystalline perfection afforded by the bulk synthesis of chiral magnetic materials has enabled scientific understanding [6-8], applications will require materials that are grown using scalable techniques such as sputtering. The chiral magnetism themself arises because of the Dzyaloshinskii-Moriya interaction (DMI) [9, 10] that is present at specimen interfaces and in the volume of noncentrosymmetric materials with broken inversion symmetry. Here we focus our investigation on thin films of cubic B20 FeGe sputtered on Si. This material lacks crystalline inversion symmetry, and has enantiomers with left-handed and right-handed crystal chiralities, which are observed by Lorentz microscopy to couple to the magnetic helicity in $Mn_{1-x}Fe_xGe$ [11]. However, in thin-film form, the magnetic state is strongly modified by substrate-induced strain[12, 13] and small grain size relative to bulk crystals.

Our work seeks to answer relevant questions including: how does sub-micron grain size and the presence of many grain boundaries alter the nanoscale chiral magnetism in this material? Imaging the noncollinear spin textures in these polycrystalline films is a direct approach to understanding questions relevant to applications.

Electron microscopy has been widely used to investigate the real space magnetization profiles of chiral and topological spin textures, e.g. helices and magnetic skyrmions [11, 14, 15]. Conventional Lorentz transmission electron microscopy (LTEM) provides nanometer spatial resolution; however, the electron beam must be defocused to obtain magnetic contrast causing strong Fresnel fringes at the grain boundaries and obscuring magnetic information [6, 15-18]. Another well-known method is off-axis electron holography, which utilizes a biprism to split the post-specimen diffracted electron beam into reference and image waves, such that at the image plane, the two waves overlap forming the electron hologram [19]. In this method, the phase shift of the electron beam can be measured and related back to the local electromagnetic field in thin samples. However, electron holography has a limited field of view, often under $5\mu m$, and the reference beam must pass through a hole in the sample. The field of view and resolution are coupled by the number of pixels on the detector, whereas for scanning transmission electron microscopy (STEM) methods the number of pixels recorded is adjustable and set by the dynamic range and noise on the scan coils – usually a few picometers in modern instruments. Holography is also sensitive to artifacts from dynamical diffraction, nonlinearities with sample thickness and small changes in crystal orientation [19-23]. In addition, all TEM methods require a thin (< 200 nm) sample, but this can be prepared from a bulk device in a few hours using a focused ion beam. For non-destructive imaging, magnetic force microscopy (MFM), a form of atomic force microscopy, utilizes a magnetic tip to scan across a magnetic sample, where the magnetic force between the tip and sample can be used to image magnetic structures such as skyrmions [24]. For MFM, resolution is limited to ~10 nanometers [25] compared to recent advances in electron microscopy in field-free mode, demonstrating sub-Angstrom spatial resolution[26].

Lorentz STEM (LSTEM) presents an alternative [27, 28] in the form of differential phase contrast (DPC) imaging, which uses a focused electron beam thus avoiding information delocalization, and has been shown to be more sensitive at measuring local magnetic fields than electron holography[29]. DPC imaging measures the deflection of a focused electron beam due to the electromagnetic field within the sample, which enables the study of the internal structure of magnetic skyrmions [30, 31]. However, there are multiple constraints to DPC imaging including: (1) limited dynamic range of the detector limits the sensitivity to detect the extremely small magnetic fields that are expected in modern devices, (2) nonlinearities introduced by the signal normalization when the detector is not perfectly centered, and (3) changing beam shape due to electron channeling, which gives rise to contrast reversals as the electron beam moves through the sample [32-34]. As a result, experiments using both LSTEM and DPC have mostly focused on single-crystal samples that exhibit sparse grains and few defects, thus presenting a structurally uniform medium for imaging.

Here, we use a version of LSTEM known as 4-dimensional Lorentz scanning transmission electron microscopy (4-D LSTEM) with an electron microscope pixel array detector (EMPAD) [35]. This technique can sense electromagnetic beam deflections by acquiring electron diffraction patterns at every x-y scan position with a high $k_x$-$k_y$ momentum resolution. From the electron diffraction pattern, a center of mass (COM) signal can be extracted and used to quantitatively reconstruct the sample's magnetic field [34]. The key advantage of the EMPAD is their high dynamic range, which enables sensitive mapping of magnetic fields, $\left(\frac{\mu T}{\sqrt{Hz}}\right)$, and captures simultaneously the crystalline and magnetic information without saturating the detector. From this, we distinguish magnetic field information from crystalline grain structure effects and probe each effect independently. In the following sections, we (1) highlight the fundamental ideas behind detection of magnetic fields with high sensitivity, (2) show that the precision relies on the number of electrons a detector can sustain, stressing the importance of a high-dynamic-range and high-speed detector, and (3) demonstrate our technical approach to disentangling magnetic signal from the signal due to grain contrast. Using this approach, we observe changes in the magnetic textures from the different DMI between adjacent grains.

## II. Deflection of the Electron Beam due to Magnetic Fields

To detect a magnetic field in Lorentz Microscopy, we consider a parallel electron beam source that is incident perpendicular to the plane of the specimen, along the z-axis. Here, we ignore stray field effects, which gives rise to an unmeasurable z-component deflection and consider only the deflections due to the projected in-plane magnetic field [36, 37]. The equation for the electron beam deflection is then given as:

$$\beta(x) = \frac{e\lambda B_0 t}{h}, \tag{1}$$

where $B_0$ is the local magnetic field, e is the electron charge, $\lambda$ is the wavelength of the electron beam, t is the sample thickness and h is Planck's constant. This deflection of the electron beam, $\beta$, can be measured by tracking the COM of the diffraction pattern as we scan the electron beam, which is related to the local magnetic field.

The argument above presents the classical treatment of the electron beam deflection in a magnetic field. On the length scale of electron microscopy, electron-specimen interactions and the beam shape should also be treated quantum mechanically. For elastic scattering, the deflection of the electron beam by the local magnetic field is the change in momentum of the electron wave function, $< \vec{p} >$ [34], which is the COM signal in our diffraction pattern. Using Ehrenfest's theorem [38], the expectation value of the Lorentz force in terms of $< \vec{p} >$ is [34],

$$\frac{d\langle\vec{p}\rangle}{dt} = \langle[\hat{H}, \hat{p}]\rangle = \langle e(\vec{E} + \vec{v} \times \vec{B})\rangle \tag{2}$$

where $\hat{H}$ is the Hamiltonian, e is the charge of the electron, $\vec{E}$ is the electric field, $\vec{v}$ is the velocity at which the electron is traveling and $\vec{B}$ the magnetic field. Again, for elastic scattering the velocity remains constant so the rate of change of momentum with time can be mapped to the rate of change of momentum with thickness into the sample. This $< \vec{p} >$ can be extracted from the intensity of the scattered wave function, $\Psi$, as:

$$\langle\vec{p}\rangle = \int \Psi^*(\vec{p})\hat{p}\Psi(\vec{p})d\vec{p} = \int \hat{p}|\Psi(\vec{p})|^2 d\vec{p} = \int \hat{p}I(\vec{p})d\vec{p} \tag{3}$$

where I is the intensity of the diffraction pattern and $\hat{p}$ is the momentum operator. From equations 2 and 3, a shift in $< \vec{p} >$ is perpendicular to the magnetic field and parallel to the electric field. Rewriting $< \vec{p} >$ in real space, we see that it is also the probability current flow, $\langle\vec{j}\rangle$, where the expectation value for $\langle\vec{j}\rangle$ is calculated for an evolving electron wave function, changing as it propagates through the specimen:

$$\langle\vec{p}\rangle = \frac{\hbar}{2i}\int\left[\Psi^*(\vec{r}, \vec{r}_p)\vec{\nabla}\Psi(\vec{r}, \vec{r}_p) - \Psi(\vec{r}, \vec{r}_p)\vec{\nabla}\Psi^*(\vec{r}, \vec{r}_p)\right]d\vec{r} = 2m\langle\vec{j}\rangle \tag{4}$$

Here, r is the real space position at an arbitrary point. $r_p$ is the position of the probe in real space, and m is the mass of the electron. Although the equations above give us a direct relation for $< \vec{p} >$ to the electromagnetic field, every signal from $< \vec{p} >$ results in a signal with contributions from the nuclear, core, valence and external electric and magnetic fields. Disentangling these contributions based on their different spatial distributions is discussed in section V. Equations (2)-(4) describe the fundamental physics used when determining electromagnetic fields from a deflection of the electron beam. To experimentally quantify the parameters that affect our signal sensitivity, we develop a framework to model and predict the sensitivity based on the parameters we can optimize in the electron microscope.

## III. Determining Magnetic Field Detection Sensitivity

The magnetic field sensitivity that we can obtain using COM imaging depends on how well we can track the deflection of the electron beam. To quantify this, we first present an analytical approach to determine the sensitivity of the magnetic field based on a model previously described by Chapman et al. for a split detector[37] (Appendix A). Here, Fig. 1 shows a schematic of a DPC detector from which the shift of scattered electron beam due to a magnetic field in the sample is detected [37].

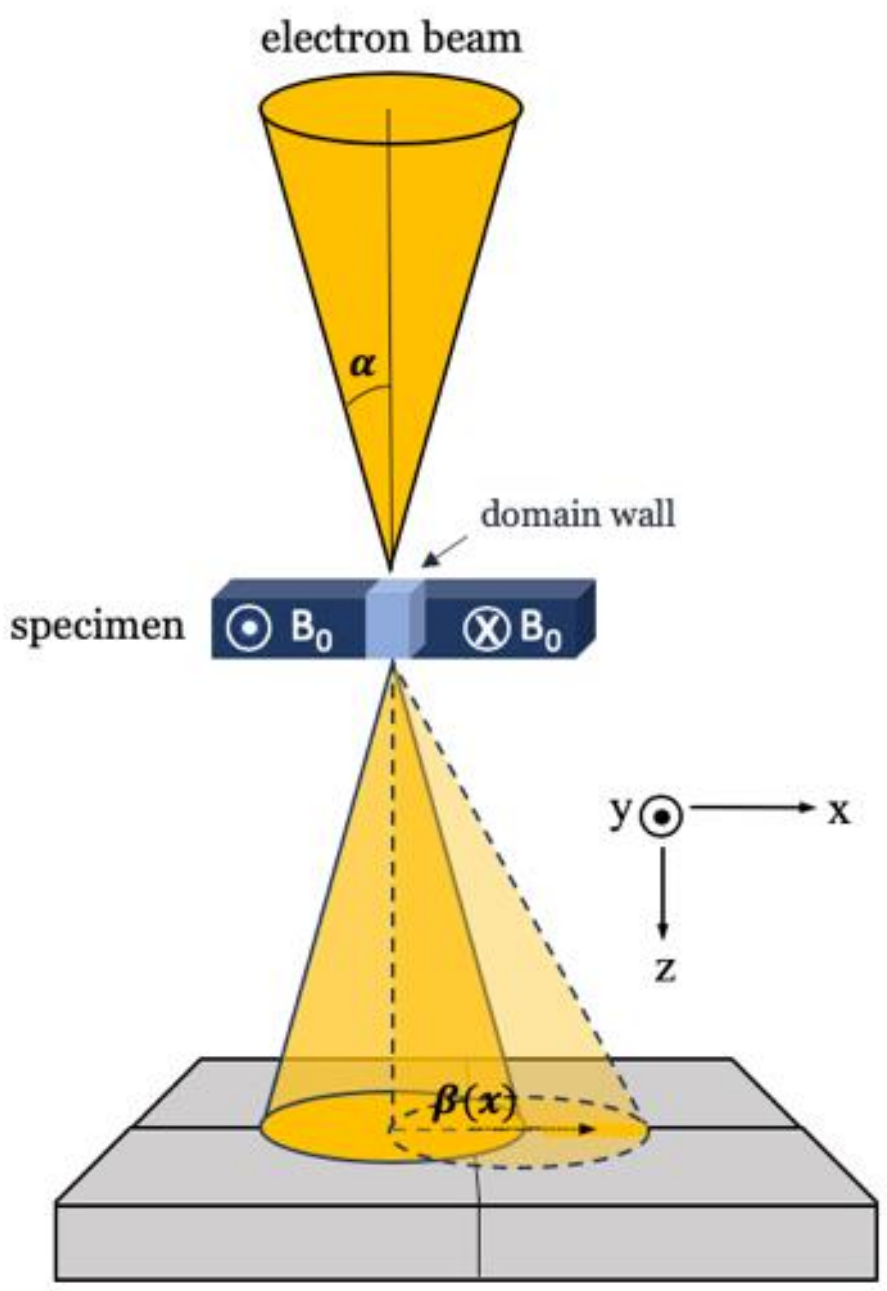

FIG. 1. Schematic of signal deflection for a DPC detector used in LSTEM, where we show change in the angular deflection of the electron beam due to the change in the Lorentz force as it is scanned across the sample from the left to right magnetic domain.

Here, the deflection sensitivity of DPC imaging has two dependencies: (1) $\alpha$, the semi-convergence angle of the electron beam, which provides the limit in spatial resolution, and (2) $n$, the number of electrons or dose of the electron beam, which tells you how many electrons are needed to sustain a signal that is adequately above Poisson noise. These two variables are parameters that can be easily adjusted in the electron microscope. Using the analytical model in Appendix A, we observe the minimum deflection $\beta_L$ as:

$$\beta_L = \frac{\pi}{4}\frac{\alpha}{\sqrt{n}} \quad (5).$$

In this analytical model, we only show how deflection sensitivity, hence field sensitivity, is dependent on a DPC detector or a detector with two pixels in one dimension. To calculate the field sensitivity of a detector with more than two pixels, we move from an analytical to a numerical model to investigate how deflection sensitivity is affected by n, the number of electrons that we use in our electron beam, and $j$, the number of pixels in our detector described in Appendix B. Doing so, we answer the question, do more pixels in a detector lead to an improved detection sensitivity to fields? We find that as a general trend, the percentage error in locating the center of mass of the beam, $(\beta_L/\alpha)$, scales as Poisson noise, $1/\sqrt{n}$ and the numerical pre-factor depends on the number of pixels.

In our numerical model, we assume a one-dimensional detector that is entirely illuminated by an electron beam represented as a top hat distribution with varying Poisson noise. We used Poisson noise to simulate the actual shape of the electron beam so that their shape is not a perfect uniform top hat probe. We chose Poisson noise because the EMPAD detector has a detector noise level of 0.007 primary e-/pixel [35], while a typical number of recorded primary electrons/pixel is ~100,000, giving a Poisson noise of $\sqrt{100{,}000} = 316$ e-. In other words, the detector noise is about 50,000x smaller than the Poisson noise and can be safely neglected. Therefore, the noise of our data is dominated by the intrinsic Poisson noise determined from the number electrons of the incident beam. Given this, we can model the distortion of the shape as only being from Poisson noise. We convert this value to a deflection and a magnetic field where the statistical model is used to simulate the sensitivity of the magnetic field measurement. For our numerical model, the detector noise is assumed to be negligible compared to the signal detected.

We vary the number of pixels in our detector, the number of electrons used in our electron beam, and hence the Poisson noise. We then calculate the COM of the electron beam on the detector over 2000 electron beam configurations at each dose setting with random varying Poisson noise to find the standard deviation of the COM measurements from the statistics of 2000 samples. The standard deviation that we obtain is the estimated random error from Poisson noise in our COM measurement, related to how precisely we can extract the signal from COM. We define a calibration factor to be $\frac{2\alpha}{(j-1)}$ in Appendix B, with $\alpha$ being the semi-convergence angle and $j$ being the number of detector pixels, so that by multiplying the standard deviation in the unit of detector pixel with the calibration factor in the unit of rad/pixel we can convert the COM standard deviation to the minimum deflection angle of our electron beam, $\beta$. In our numerical model, we choose $\alpha$ to be 200 μrad because it corresponds to a spatial resolution of 7.6 nanometers for a 200 kV electron beam, typical of the resolution needed to image magnetic skyrmions in single crystal FeGe. Because we assume a square electron beam shape for simplification in the simulation, the associated minimum detectable deflection angle for a split detector is $\beta_L = \frac{\alpha}{\sqrt{n}}$ (Appendix B). This is a different numerical prefactor from the round detector case of equation 5. Our simulation results in the case of a two-pixel detector agree with the minimum deflection angles calculated from the analytical model with a split detector in the version of square probe shape. From Fig. 2(a), we find that our deflection, $\beta$, is strongly dependent on the total number of electrons, and only weakly on the number of pixels in our detector, with diminishing returns once more than 10 pixels per disk are added.

To convert $\beta$ to magnetic field, we use a beam energy of 200 keV and a sample thickness of 100 nm [Fig. 2(b)]. Here, our results show that if we want to observe magnetic fields with strength less than 1 milliTesla (mT), the number of electrons must be greater than 1 million. From Fig. 2(a) and 2(b), we find that a detector with high dynamic range is more important for signal sensitivity than a detector with more pixels. Traditional arguments for detectors with many pixels emphasize the importance of angular resolution, where more pixels in the detector would give higher sensitivity in deflection detection. Here, we show through our models that this is not necessarily the case for COM measurements. We observe that even with larger number of pixels, deflection sensitivity is still limited if a detector cannot collect or sustain a high number of electrons integrated over all channels. While there is diminishing return in improving the SNR with an increased number of pixels, for detectors with poor dynamic range, there sometimes might be a benefit to spreading the total dose over more pixels to avoid saturation, provided the dose/pixel is much larger than the readout noise per pixel. In summary, we find that we do not need many pixels on our detector to detect small fields. Instead, the most important thing is the ability of a detector to handle large total doses; here, dose/pixel becomes important. For detecting very small deflections or fields, dynamic range is the limiting factor.

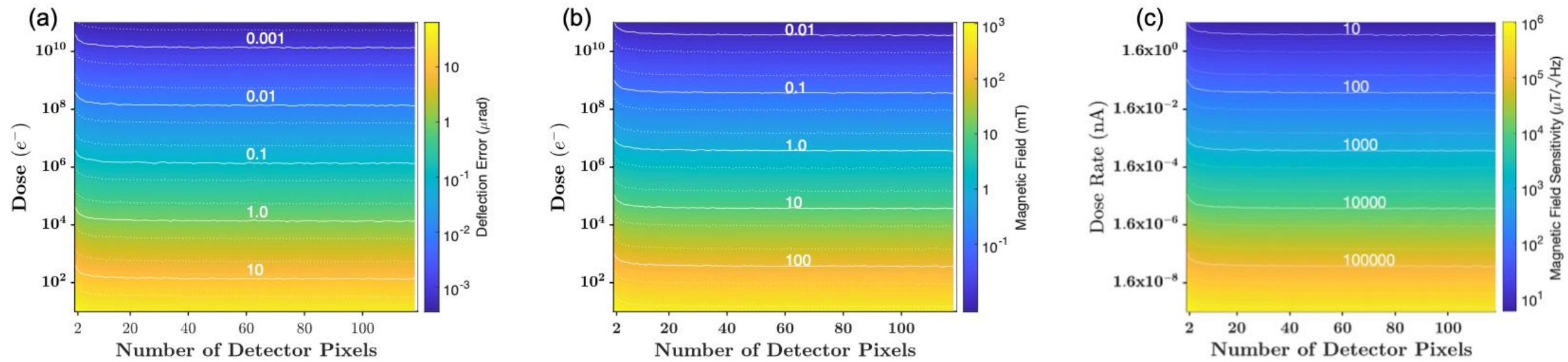

FIG. 2. Numerical simulations using a 200 μrad semi-convergence angle at a beam energy of 200 keV, and sample thickness of 100 nm. We show (a) the minimum measurable deflection angle, (b) the smallest measurable magnetic field, as a function of electron dose and number of detector pixels, and (c) the dependence of magnetic field sensitivity ($\frac{\mu T}{\sqrt{Hz}}$) on beam current (nA) and number of detector pixels. The white lines represent the contours for deflection error (a), magnetic field (b) and B field sensitivity (c) as 0.2, 0.5 (dotted lines) and 1 (solid lines) times powers of 10. From our plots, we observe that field resolution is more dependent on the number of electrons than the number of pixels on a detector, with diminishing returns beyond 10 pixels.

To measure these small deflections with high precision experimentally, we need cameras with higher dynamic ranges and faster read-out speeds than what is possible with scintillator-based technologies. The EMPAD presents one such a solution [35]. Compared to traditional diffraction detectors such as CCDs, the EMPAD has a relatively fast acquisition at 0.86 ms per frame and a high dynamic range of 30 bits; this means that the EMPAD can detect one to one million electrons without detector saturation at high signal to noise. *Using the EMPAD, DC magnetometry using a COM approach can achieve a comparable sensitivity, albeit at higher spatial resolution, to other high-sensitivity DC magnetic imaging techniques, such as scanned diamond nitrogen-vacancy centers that achieve sensitivities in the few* $\frac{\mu T}{\sqrt{Hz}}$ *range using optically-detected magnetic resonances[39-41]. AC modulation can improve the sensitivity in all cases.* Field sensitivity as a function of beam current is shown in Fig.2c. For LSTEM, the beam deflection comes from sample in-plane magnetic field summed over projection, and the field sensitivity scales inversely with square root of electron beam current and inversely proportional with sample thickness for thin samples (Appendix A). Sample thicknesses greater than an electron mean free path will suffer from electron multiple scattering out of the central beam which would exponentially attenuate the deflection signal, and the tradeoff between the two effects implies an optimal range of sample thickness. Peak field sensitivity, ~ 10 $\frac{\mu\mathrm{T}}{\sqrt{\mathrm{Hz}}}$, can be obtained at a film thickness of around 100 nm, and dose rate limit of 10nA, which is a high, but still-usable beam current for a cold field emission source. Here, using simulations, we have analyzed how to optimize the detection of extremely small magnetic field variations in the electron microscope, highlighting the importance of a high-dynamic range diffraction detector such as the EMPAD.

## IV. Experimental Results

In this section, we apply 4-D LSTEM to study bulk single-crystal FeGe, a well-characterized material system [6, 15, 42-44], before proceeding to our investigation of polycrystalline materials in section V. The quantitative comparison to a known material allows us to test the accuracy and precision of the LSTEM measurements. The sample preparation of single crystal FeGe was discussed previously in [45-47] . Samples are prepared using a focused ion beam and subsequently imaged at 300 keV with a 230 micro radian semi-convergence angle on a FEI-Titan. The beam current for imaging is 198pA and dwell time is 1ms. The sample was cooled to 240 K using a Gatan cryo-holder and imaged under at 130 mT field perpendicular to the sample plane. The objective lens was used to induce the 130 mT field to the sample.

We calibrate the strength of the objective lens using Hall probe measurements, with a Hall probe mounted in a Protochips Aduro holder with electrical feedthroughs. As the strength of the lens changes, we can relate their strength to a value for the induced magnetic field on our Hall probe.

We perform 4-D LSTEM on single crystal B20 FeGe, a sample with a uniform background. During scanning we record the full convergent beam electron diffraction (CBED) pattern at high-dynamic range and high speed [Fig. 3(a)]. The CBED pattern contains all of the magnetic and crystallographic information, including the Lorentz force, which is extracted from the beam COM. The sample's magnetic induction field for the x and y direction are shown in Fig. 3(b) and (c), respectively. Using Fig. 3(b) and (c), we reconstruct the magnetic field magnitude, which shows the skyrmion lattice [Fig. 3(d)]. Our in-focus 4-D LSTEM image of the skyrmion lattice agrees quantitatively with prior measurements using LTEM[15]and DPC [30]. To account for the thickness of the sample in our magnetic field measurements, we use electron energy loss spectrum (EELS) to map the inelastic mean free path, and from this we estimate the thickness to be about 100nm. We confirmed these estimates locally from the fractional intensity remaining in the central beam of each EMPAD pattern using the Beer-Lambert law with an analytic estimate of the elastic mean free path[48], and the two measurements agree to within 10%. However, when the sample becomes too thick (larger than a mean-free path, or strongly diffracting), the intensity of the incident beam is reduced and multiple scattering and dynamical diffraction dominates the signal. As a practical matter, the phase approximation breaks down at about a mean free path (~120-150 nm) and it becomes difficult to distinguish separate the Lorentz deflection from diffraction artifacts.

To estimate the noise in our measurements, we take a line profile of the magnetic field in Fig. 3(c) and perform a fit to our line. Here, we obtain the root mean square (RMS) error by measuring the deviations in the line profile of the magnetic field compared to our fit [Fig. 3(e)]. We find a root mean squared (RMS) error of 3.6 mT and observe that we indeed have mT sensitivity in field.

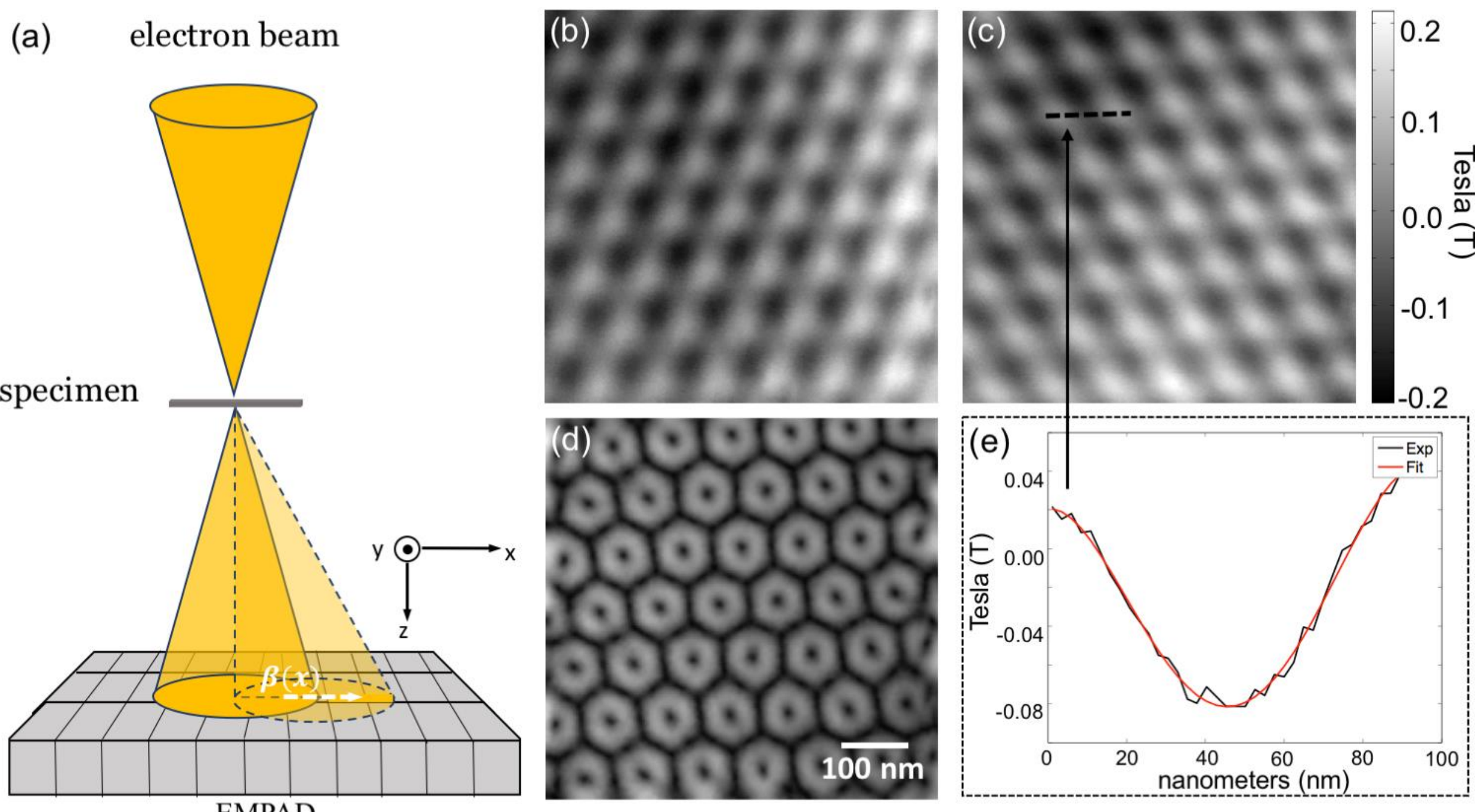

FIG. 3. 4D Lorentz-STEM of single crystal B20 FeGe imaged at 240K with 130 mT applied magnetic field on a FEI-Titan. (a) Schematic of magnetic imaging with EMPAD in Lorentz-STEM mode. We capture the full scattered distribution in momentum space with the EMPAD at every scan position, from which we can also reconstruct the magnetic field components in the (b) x and (c) y directions (in units of Tesla). From

(b) and (c), we obtain (d) the magnitude of the skyrmion field. We record a line profile from the magnetic field deflection in y (c), and we compare it to their fit (e). Here, we observed a root mean square (RMS) noise of 3.6 mT. White scale bar in (d) also applies to panels (b) and (c).

Aside from magnetic fields, the 4-D LSTEM with the EMPAD also simultaneously provides the information needed to generate images that are usually obtained from conventional techniques such as annular dark field, bright field, and specialized techniques such as quantitative measurements of thickness, strain, tilt, polarity, atomic fields, and the long-range electric fields. Although we can detect sensitive magnetic fields, crystallographic scattering effects could dominate signals from real specimens, leading to artifacts or misinterpretation of magnetic signals. For spintronic devices, magnetic specimens are typically sputtered as thin films where grain sizes range from a couple of nanometers to microns. One such material is sputter-deposited B20 FeGe on Si [3, 12, 39, 40, 49]. Probing the magnetism of sputtered B20 FeGe is extremely challenging because the sub-micron crystal grains exist on the same length scale as the characteristic length scale of the magnetic texture. Therefore, disentangling magnetic effects from grain contrast is extremely important. We present an approach to disentangle the signal of magnetic contrast from grain contrast, allowing each to be probed independently. This is enabled by the high dynamic range of the EMPAD, where multiple imaging modes can be used to extract and separate the two signals.

## V. Disentangling grain and magnetic contrast

As a starting point, we must first consider how electron scattering is affected when both magnetic and grain contrast are present. Previously, Chapman et al. [50] showed that by using an annular quadrant detector instead of a solid quadrant detector, the unwanted grain contrast could be reduced, and a smoother magnetic contrast could be obtained by narrowing the collection angles of the detector. Kohl et al. [51] discussed how using only a narrow range of annular cut-off angles relative to the direct electron beam acts as a low-pass filter for the image. Krajnak et al. [52] showed that magnetic contrast can also be enhanced by tracking only the edge of the bright field disk, which also serves to suppress high-spatial-frequency, non-magnetic features in very fine grain materials. This can be understood in terms of Kohl and Majerts's [51] analysis by recognizing that magnetic field signals are enhanced due to their slowly-varying long-range potentials. Diffraction contrast from grains and grain boundaries, on the other hand, are encoded as short-range potentials, where this signal can be smaller than the probe size for a large electron beam [53]. Cao et al. [53] explored the asymptotic limits for both long-range potentials where the potential varies slowly across the probe shape, and short-range potentials where the potential is much smaller than the probe size, showing how their signals are distributed differently in momentum space. For magnetic imaging, where the probe is a few nanometers in size, magnetic fields are long-range with respect to the probe, and changes in grain structure and contrast are short range. A short-range potential, such as the atomic arrangement at a grain boundary, changes intensity distribution inside the bright field disk and do not deflect the disk boundaries. A long-range potential, such as a magnetic field in a ferromagnet, uniformly shifts the entire bright field disk similar to the classical analysis of Ref. [36]. These different angular distributions are difficult to track with only a quadrant detector. A key aspect of this angular-separation technique is having a sufficient number of pixels on the detector to separate signals from long and short-range potentials, while recording sufficient signal from diffracted beams to track grain orientation simultaneously.

Recent work from Wang et al. [54] has taken a different approach to enhancing magnetic contrast in polycrystalline FeGe by noting that equation 5 shows angular precision scales linearly with the convergence angle. Thus higher sensitivity can be reached by choosing the smallest aperture (and hence the largest probe) that can just resolve the magnetic features (~20 nm). However, at this limit, the magnetic features are no longer much larger than the probe size, and instead of appearing as shifts of the diffraction disk, the magnetic contrast is now present as features inside the disk themself. Consequently, disk-edge tracking is no longer effective for separating magnetic and grain contrast under those conditions. Instead bandpass filtering becomes necessary.

To investigate long and short-range potentials and their effects on magnetism, we study B20 FeGe (176 nm thickness) films sputtered on Si to test how the intensity distribution signal changes in the diffraction pattern using the full 128x128 pixels on the EMPAD and their high dynamic range. We polished our sample in plan-view using a 3 degrees polishing angle to remove the Si region at the tip of the sample, exposing a thin wedge of free-standing FeGe. We use a FEI-Tecnai F20 at 200 keV with a Gatan double-tilt cryo-cooling holder to image our sample at 100 K in 4-D LSTEM mode with the EMPAD, where we collect a diffraction pattern at each scan position. We do not use the main objective lens for focusing because it induces a 2 T field to the sample, which would fully saturate their magnetization and erase all chiral magnetic texture. Instead, we use the condenser lenses to focus the probe and we use the objective and condenser mini lenses to either null the field on the sample or add a small external vertical field. In this experiment, we chose a semi-convergence angle of 615 μrad giving us spatial resolution of 2.5 nm.

Here, we can explore the short and long-range potentials as intensity distributions in our diffraction pattern, where we can enhance the signal for magnetic fields [Fig. 4(i)] and decouple it from the grain structures [Fig. 4(j)]. To decouple the magnetic field from grain contrast, 1) we vary a virtual aperture on our diffraction pattern and 2) for each aperture, we extract a COM signal, where the sampled angular distribution of the scattered electron beam is limited by the aperture sizes [Fig. 4(a)-(c)]. Following Cao et al.'s [53] analysis, when the size of the aperture is less than the size of the incident diffracted disk, we treat the COM signal as coming from only the short-range potentials such as grain boundaries in the sample. When the aperture is larger than the size of the incident beam, we treat the signal as coming from both the long-range field (magnetic field) and the diffracted beams (grain structure). Here, we choose virtual collection aperture sizes of 430 μrad, 700 μrad and 4.6 mrad [Fig. 4(a)–(c)], which correspond to the radii of the apertures. At aperture sizes smaller than our semi-convergence angle, 615 μrad [Fig. 4(a)], we find that the signal only comes from the short-range potentials, i.e. the grain contrast [Fig. 4(e)]. When we increase our aperture size to 700 μrad [Fig. 4(b)], which is slightly larger than our semi-convergence angle, we observe a signal from the magnetic field in the form of magnetic helices [Fig. 4(f)]. When we extend our aperture size further to 4.6 mrad [Fig. 4(c)], we observe both the signal for the magnetic field and grain contrast [Fig. 4(g)], similar to Fig. 4(f); however, there is an increased background from thermal diffuse scattering (including Kikuchi bands and multiple scattering) at higher scattering angles, as well as contrast modulations from Bragg beams, reducing the signal to background ratio of the magnetic field signal than in Fig. 4(f). From this, we see the that optimal outer angle for collecting the magnetic signal comes from choosing an aperture angle that is slightly larger than our semi-convergence angle.

To suppress the short-ranged grain boundary and diffraction contrast, we exclude the interior of central disk from the analysis. Instead, we focus on the edge shift by evaluating a narrow annulus around the disk edge by forming Fig. 4(h), which is the COM image when the aperture is slightly smaller than the incident disk [Fig. 4(e)] subtracted away from the COM image when the aperture is slightly larger than the incident disk [Fig. 4(f)]; schematic shown in Fig. 4(d). From the difference signal in Fig. 4(h), we can see that magnetic contributions are enhanced, and grain contrast has been significantly reduced compared to Fig.4(e-g). It is worth pointing out that dynamical scattering from a relatively thick crystalline sample also leads to intensity variations in the center disk, which can also be significantly reduced by the subtraction scheme of Fig. 4(h). Therefore, our approach also improves field measurements for thicker samples because they reduce distortions from strong dynamical scattering. Finally, we use the COM-X and COM-Y from the decoupled magnetic contrast and reconstruct a vector map showing magnetic helices [Fig. 4(j)], where we can plot the amplitude and direction relative to the magnetic induction color wheel, shown as the inset of Fig. 4(j).

On the other hand, we can investigate the grain contrast by suppressing the magnetic signal; here, we exclude the edge of the central disk by subtracting Fig. 4(f) from Fig. 4(g) such that we subtract away magnetic effects and leave mostly grain contributions. The resulting COM signal reflect the tilt of the Ewald sphere and shifts of the Kikuchi bands, giving a qualitative sense of changes in grain orientation with reduced sensitivity to thickness changes [Fig. 4(j)]. When we compare Fig. 4 (i) and (j), we find magnetic helical phases in Fig. 4(i) do not appear in our grain contrast image [Fig. 4(j)], where most of the magnetization components have been subtracted away. For finer grain sizes and overlapping grains, our technique would still work so long as we can make our semi-convergence angle small enough to separate the center disk from the rest of the diffracted disks. When comparing the Fig. 4(i) and Fig. 4(j), we find changes due to the electrostatic potential; here, the yellow contrast in Fig. 4(i) comes from the thickness gradient of a wedge sample, which is a long-range electrostatic contribution to the Lorentz signal. In addition, charging artifacts such as two small pieces left over from polishing can be seen on the right of Fig. 4(i), sticking out from the edge of the sample. It is not present in the orientation map of Fig. 4(j) which was designed to reduce contributions from slowly varying thickness changes.

Having established a method to resolve nanoscale magnetic contrast in the presence of structural disorder, we now discuss our observations of magnetic helices. The orientation of helical magnetic states can be defined in terms of a Q-vector – the vector about which the magnetic moment spirals as a function of position. Thin-film FeGe (and other B20 magnetic materials) samples typically exhibit a helical state with a Q-vector oriented out-of-plane, which has been verified by both neutron scattering experiments [13] and by the ferromagnetic resonance of the helical state [12]. This tendency can be attributed to substrate-induced tensile strain in the film which produces anisotropy oriented in the sample plane [12, 30, 55-58], consistent with strain-dependent experiments in single-crystal FeGe samples [56]. To see helices in an electron microscopy experiment, the Q-vector must lie in the plane. In the case of out-of-plane Q-vectors, the magnetization always remains in the sample plane as it spirals about the out-of-plane direction. Therefore, any deflection acquired by the electron beam as it passes through the upper part of the film will be compensated by a deflection in the opposite direction in the lower part of the film. In contrast, when the Q-vector lies in the plane, the beam deflections are not canceled, and they vary with the lateral position leading to the stripe-like pattern seen in Fig. 4. This both offers an opportunity to study the helical and skyrmion textures as they thread through different crystalline grains. These topics are addressed in Sec. VI.

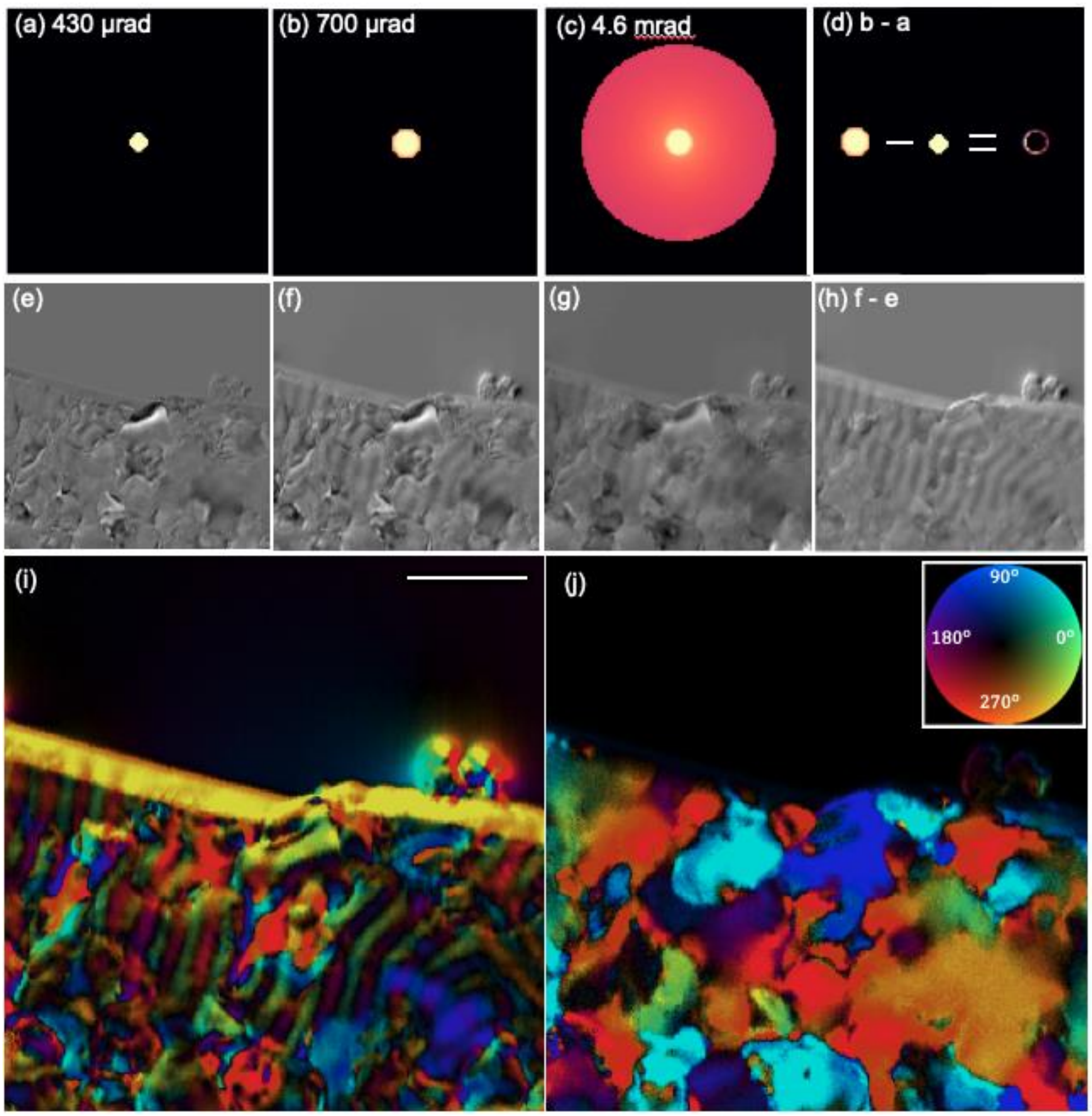

**FIG. 4. 4D Lorentz-STEM of polycrystalline B20 FeGe thin film on silicon substrate imaged at 100K with 0 mT applied magnetic field on a FEI-F20.** Here, we optimize magnetic field contrast by varying the collection aperture size. With a 615 μrad convergence semi-angle, we calculate the COM in the x-direction, using with virtual collection aperture sizes of (a) 430 μrad (b) 700 μrad and (c) 4.6 mrad. Figures (e) – (g) show the x-component of the COM images corresponding to apertures (a) – (c). In (d), we use our subtraction technique where we subtract the COM image when the aperture is slightly smaller than the bright field disk (a) from when the aperture is slightly larger than the bright field disk (b) so that we can observe only shifts from the edge of the bright field disk (d). Here, (h) corresponds to the COM image by subtracting (e) from (f). By exploiting the long-range potential due to the magnetic field through subtraction, we can extract both COM-X and COM-Y and turn them into vector components of the magnetic field, where we observed that grain contrast contributions have been greatly reduced. We also perform our subtraction technique to recover only the grain contrast by subtracting (f) from (g), where we take their COM-X and COM-Y to create a false colored vector map in (j), which shows only grain contrast image (as a tilt of the Ewald Sphere). The color wheel plots the phase of the field or crystal orientation as the hue in a continuous color scale, and the COM amplitude is mapped to the brightness. We used COM-X and COM-Y as vector components and represent them as an amplitude and phase, such that a change in direction of magnetic induction [Fig. 4(i)] or grain orientation [Fig. 4(j)] can be shown. White scale bar in (i) is 200 nm.

## VI. Skyrmions and Helical Magnetic Phases at Grain Boundaries

The ability to image structural features and magnetic effects at the same time has enabled us to study how skyrmions and helical phases change along the boundaries of individual FeGe grains. Previous studies in crystalline domains of alloyed $Mn_{1-x}Fe_xGe$ (x~0.7) helimagnets using Lorentz-TEM have shown correlations between skyrmion helicity and crystal chirality with varying chemical compositions [11]. For chiral magnets made from thin-film sputtering with smaller polycrystalline granules, grain contrast and substrate effects have limited direct imaging of these chiral magnets due to image distortions caused by Fresnel fringes in Lorentz-TEM, where artifacts arising from grain contrast could be misinterpreted for magnetic effects [16-18, 52]. In addition, structural effects from the grains in B20 FeGe were previously shown by electron back scattering diffraction (EBSD) to be twinned, which causes switching of crystal chirality [12], adding another layer of difficulty when decoupling magnetic from grain effects. Therefore, to decouple all of these effects and observe the intrinsic magnetic properties of the B20 FeGe magnets, we utilize 4-D LSTEM with the EMPAD.

To investigate the relationship between magnetic textures and crystal chirality of sputtered B20 FeGe on Si, we prepared a planview sample using a focused ion beam. The sample is thinned at ~ 3° with respect to the planar surface to create a gradient of Si thicknesses while milling, this also creates a thin wedge of free-standing FeGe thin film in regions where the Si is completely milled away. To account for the thickness of the sample in our magnetic field measurements, we use electron energy loss spectroscopy (EELS) to measure the thickness in terms of inelastic mean free paths [48]. From our results, our region of interest [Fig. 5] is measured to be about 130nm. We then performed 4D Lorentz-STEM at 100K with a 340 $\mu$rad semi-convergence angle and 111 mT applied magnetic field on a Thermo Fisher Titan Themis at 300kV. To decouple magnetic and grain effects, we utilize the technique described in section V. In this section, we will describe our observations of skyrmion and helical phases in adjacent grains with opposite and similar crystal chirality.

We start by looking at how the skyrmion helicity is affected in two adjacent grains with different crystal chirality [Fig. 5]. Here, we find that the top grain shows anti-clockwise rotations of skyrmions and the bottom, clockwise rotations [Fig. 5(a)]. We examine the origin of these skyrmion rotations using annular dark field [15] where diffraction contrast from the contour of the two adjacent grains marks the grain boundary, shown as a black dashed line [Fig. 5(b)]. Using an atomic sized probe, we obtain atomic resolution images of the top and bottom grains (Fig. 5 (c-d)) at room temperature. The two grains are viewed in projection down the (111) zone axis, with corresponding simulations in Fig. 5 (c-d). Structurally, the differences in chirality can be seen in the tiny rotations of the purple atoms (Ge) that have clockwise rotations for the top grain (Fig. 5(c)) relative to the corner orange atoms (Fe and Ge in projection) and anticlockwise rotations for the bottom grain (Fig. 5 (d)). When compared to the simulations for right and left-handed crystal chirality as inset of Fig. 5 c and d, the top grain exhibits right-handed crystal chirality (Fig. 5(c)) and the bottom left-handed (Fig. 5(d)). This result was further confirmed by Kikuchi bands from the top (Fig. 5(e)) and bottom (Fig. 5(f)) grains where the positions of the wide bands (marked by the yellow arrows) and thin bands (marked by the blue arrows) are switched across the grain boundary. From these results, we infer as in Ref [11], that magnetic helicity is coupled to the crystal chirality.

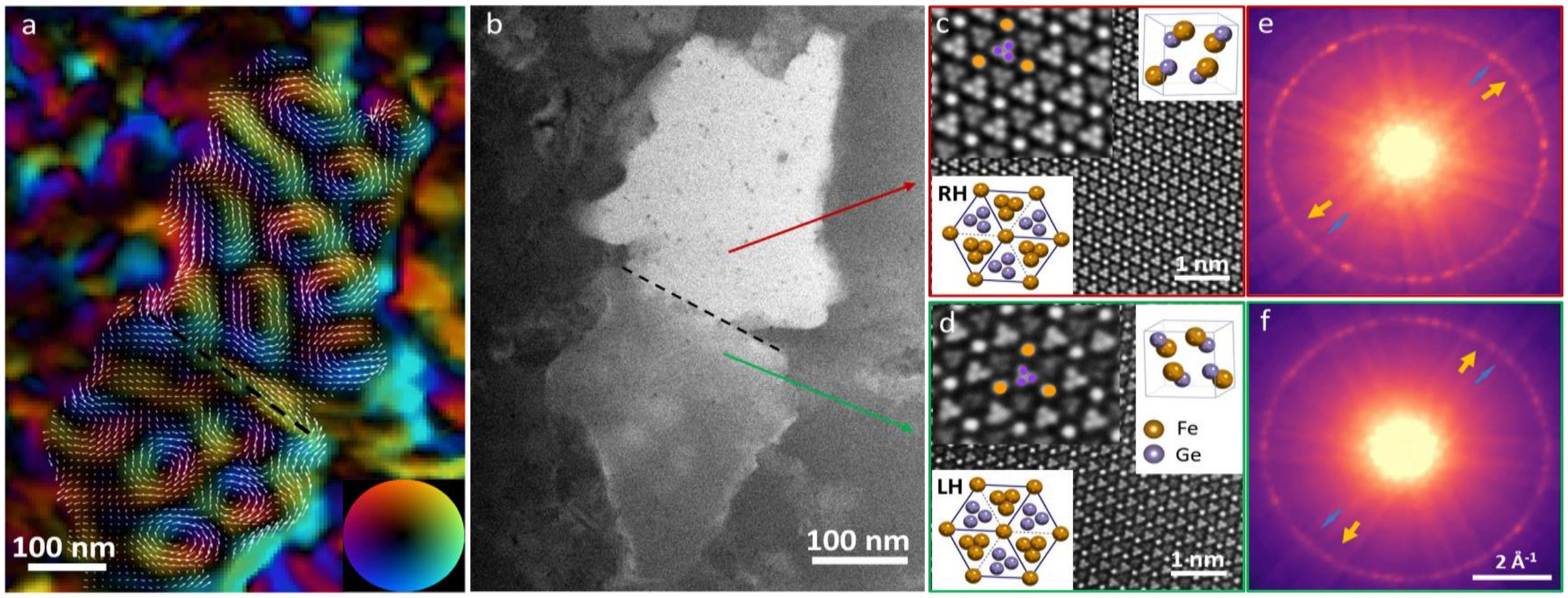

**Fig. 5. The Skyrmion helicity is coupled to the crystal chirality of B20-FeGe grains.** (a) 4D Lorentz-STEM showing skyrmion helicity inverts on two adjacent grains, measured at 100K with 111 mT applied magnetic field on a Thermo Fisher Titan Themis at 300kV. The same technique for plotting the vector fields of Fig. 4 is used here. The arrows indicate the direction of the local magnetic field. There is a helicity inversion across the grain boundary, as the top grain shows anticlockwise skyrmion vortices and the bottom shows clockwise vortices. (b) Annular dark field image shows contrast from the two adjacent grains with a grain boundary marked by a black dashed line. (c)& (d) are atomic resolution HAADF images with (c) from the top grain and (d) from the bottom grain viewed down the [111] zone axis, showing the reversal of crystal chirality, as illustrated by the colored atoms in the insets. The change in crystal chirality can be tracked by the tiny rotation of the marked purple atoms(Ge) relative to the corner yellow atoms(Fe and Ge in projection), where it's clockwise in (c) and anticlockwise in (d). This is also reflected in the Kikuchi bands of the convergent beam diffraction patterns of (e) and (f) taken from the top and bottom grains respectively. The positions of the wide bands(marked by yellow arrows) and thin bands(marked by blue arrows) switch between the 2 grains, consistent with the change in crystal chirality.

To test our inference, we study helical phases and skyrmions when adjacent grains have the same crystal chirality. Using the same technique described above, we find that the magnetic helical phases remained unchanged as they cross between two adjacent grains under no applied magnetic field [Fig. 6(a)]. When we turned on the magnetic field to 111 mT, skyrmion rotations in both grains exhibit anti-clockwise rotations even at the grain boundary [Fig. 6(b)]. Here, annular dark field image shows contrast from two adjacent grains with a grain boundary marked by a dashed line [Fig. 6(c)]. At atomic resolution, annular dark field image reveal that the twin boundary of both grains is projected down the [111] zone axis, where crystal chirality remains the same across the grain boundary, albeit a 180 degree in-plane rotation [Fig. 6(d)], highlighted as insets of Fig. 6(d) exhibiting right-handed crystal chirality. This is also consistent with the related Kikuchi diffraction patterns of Fig. 6 (e) and (f) taken from the top and bottom grains respectively. Here, the positions of the wide (yellow arrows) and thin (blue arrows) Kikuchi bands are the same across the grain boundary, showing that crystal chirality remains unchanged. From our two experimental results (Fig. 5 and 6), we believe that magnetism is coupled to the crystal chirality, such that the orientations of skyrmions and helical phases change with respect to the local crystal chirality.

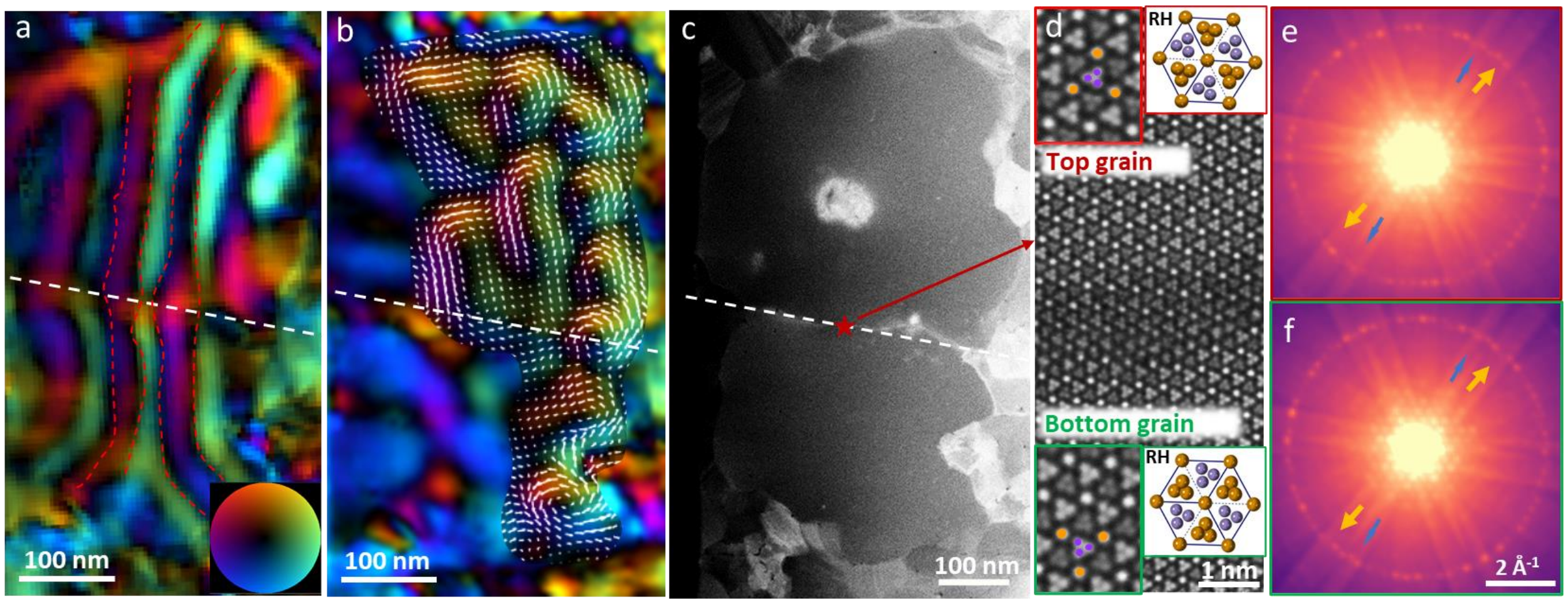

**Fig. 6. Skyrmion helicity is the same when adjacent grains have the same crystal chirality:** (a) &(b) 4D Lorentz-STEM showing the magnetic helicity remains unchanged for two adjacent grains in both (a) the helical phase with no applied magnetic field, and (b) the skyrmion phase with 111 mT applied magnetic field, measured on a Thermo Fisher Titan Themis at 300kV. Same technique for plotting vector field of Fig. 4 is used here. The helicity stays the same across the grain boundary. (c) Annular dark field image shows contrast from two adjacent grains with a grain boundary marked by a dashed line. (d) is an atomic resolution HAADF image right from the twin boundary with both grains viewed down the [111] zone axis showing the crystal chirality stays the same across the grain boundary. The zoomed-in insets of the images show both crystal structures match with right-handed FeGe, other than a 180 degree in-plane rotation. It's also consistent with the related diffraction patterns of (e) and (f) taken from the top and bottom grains respectively. The positions of the wide (yellow arrows) and thin (blue arrows) Kikuchi bands are the same across the grain boundary, again showing the crystal chirality is unchanged.

To further confirm our results, we investigate the effects of grains on the helical configuration by performing micromagnetic simulations with Mumax3 [59]. We first randomly generate grains with 320 nm average size using Voronoi tessellation [59] [Fig. 7(a)]. Next, we randomly assign either positive or negative DMI coefficients to these grains as shown in Fig. 7(b). From magnetometry measurements, we confirm that our films have a small easy-plane uniaxial anisotropy, $K_u$, approximately −3500 J/m$^3$ [6, 12]. We note that the $K_u$ of the sample may be altered in the process of preparing it for 4-D LSTEM due to mechanical polishing and thickness variation. To account for this effect, we vary $K_u$ between 0 and −3000 J/m$^3$ in our simulations and obtained a very similar spin configuration. In Fig. 7, we show the simulation for $K_u$ = -3000 J/m$^3$. Particularly, we start the simulation by initializing the system into a random magnetic configuration and allow it to micromagnetically relax to their ground state. The spin configuration of this ground state is shown in Fig. 7(c) as color-coded for the magnetization direction. When we applied at 0.3T field to the white dashed region in Fig. 7(c), which has been magnified in Fig. 7(d), we observed skyrmions with clockwise and anticlockwise rotations at the grain boundary. The applied magnetic field is larger than the one used in the experiment, because our micromagnetic simulations are performed at zero temperature; therefore, the skyrmion phase shifts to higher magnetic fields. Using micromagnetic simulations, we noticed that the helical vectors shift and the skyrmion rotation directions change at the grain boundaries, where we observed both in-plane and out-of-plane components of the helical phases [Fig. 7(c)]. In our experiments, only in-plane component of the spin configurations can be detected using shifts of the electron beam. Although the out-of-plane component cannot be detected experimentally, unless we tilt the sample, we find in both simulation and experiments that there are helical vector shifts and skyrmion rotation direction changes at the grain boundaries, suggesting that the sign of the DMI is coupled to crystal orientation. We conclude that for FeGe on Si(111), the crystal grain orientation couples to the crystal chirality, and thus to the DMI.

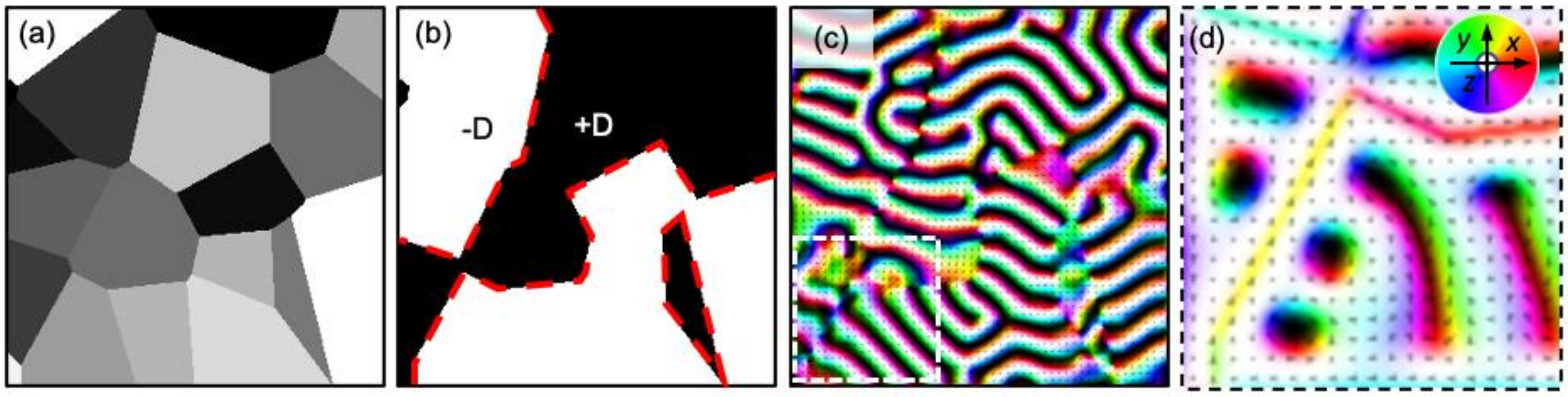

FIG. 7. Micromagnetic simulations of disordered medium of FeGe thin films. (a) Randomly generated grains, in which the opposite sign of DMI coefficients are assigned (b). (c) The relaxed ground state spin texture, where spatial shift of the helical vector is observed. (d) Magnified region white dashed region in (c), where we applied a 0.3T field and observed skyrmions with clockwise and anticlockwise rotations at the grain boundary. The simulation window is 1024 x 1024 $nm^2$. Color wheel in (c) shows the orientation of the magnetization where there is both in-plane and out-of-plane magnetic components.

## **VII.** Conclusion

In this paper, we provide three themes: (1) we quantify the sensitivity of an electron beam deflection in LSTEM and their dependence on the number of electrons and pixels on the detector, (2) we find having a high dynamic range detector is essential when measuring small electromagnetic fields, and (3) we show how we can effectively use all the scattering information from the CBED pattern collected by the EMPAD to disentangle magnetic and grain contrasts from each other. Decoupling them together enables an approach for efficient Lorentz imaging of magnetic samples simultaneously with grain structure, and investigations of changes in the magnetic field from crystal structures at the nanometer scale. We believe that our approach is important in the characterization of real devices where small changes in grain structure can be critical to device performance.

More systematically, we find that the sensitivity in magnetic field detection is more dependent on the number of electron than on the number of pixels in our detector, where high dynamic range detectors are a necessity when imaging μT-level magnetic fields. Whereas, having more pixels in our detector becomes important when we decouple signals from magnetic and grain contrast. Here, we filter long-range (magnetic contrast) and short-range (grain contrast) potentials using our acquired CBED pattern for thin film sputtered B20 FeGe on Si. We observed helicity changes in skyrmions where grain structures are of opposite crystal chirality and no change in helical phases and skyrmion rotations when the crystal chirality is the same, even when there is a 180 degree in-plane rotation of the crystal structure. By performing micromagnetic simulations, we generate random grain orientations and signs of DMI. From these simulations, we observed that the DMI is coupled to the crystal chirality, which can dictate how the skyrmion rotation and helical vectors change when going in between grains. Our technique can also be applied to other chiral magnets. Here, understanding the nanoscale physical and magnetic properties of chiral magnetic materials is key to advancing next generation magnetically-driven devices.

**ACKNOWLEDGMENTS:** Electron microscopy experiments of thin-film FeGe samples and equipment were supported by the Cornell Center for Materials Research, through the National Science Foundation

MRSEC program, award DMR 1719875. Electron microscopy measurements of single-crystal FeGe samples were supported by DARPA under cooperative agreement D18AC00009. Thin-film FeGe growth and micromagnetic modeling were supported by the Department of Energy Office of Science under grant No. DE-SC0012245. Growth of single-crystal FeGe was made by M.J.S. and S. J., who are supported by NSF grant ECCS-1609585. M.J.S. also acknowledges support from the NSF Graduate Research Fellowship Program grant number DGE-1256259.

**Appendix A**: Determining Magnetic Field Detection Sensitivity: Analytical Model

How sensitive the measurement for magnetic field using center of mass (COM) imaging depends on how well we can track the deflection of the electron beam. To quantify this, we follow an analytical approach to determine the accuracy of the magnetic field using a model previously described by Chapman et al. [27, 60]. In this analytical model, the deflection sensitivity of our signal has two dependencies: (1) $\alpha$, the semi-convergence angle of the electron beam, which provides the limit in spatial resolution, and (2) n, the number of electrons or dose of the electron beam, which corresponds to the number of electrons needed to sustain a signal that is adequately above Poisson noise. From Chapman et al., assuming a round detector divided into quadrants, the signal S is the difference in intensity between the left and right halves of the detector, described as:

$$S = \frac{4\beta_L}{\pi\alpha} \quad \text{(A1)},$$

where $\beta_L$ corresponds to the deflection angle of the electron beam from the local magnetic field of the sample. Here, the dose n is

$$n = \frac{B\pi^2\alpha^2D^2\tau}{4e} \quad \text{(A2)},$$

where $e$ is the electron charge constant, B is the brightness of the electron gun, D is the beam diameter, and $\tau$ is the acquisition time. The signal with respect to the dose is $S*$n, and the noise due to Poisson statistics is $\sqrt{n}$. We calculate the signal to noise ratio (SNR) as $s = S * \sqrt{n}$. To estimate the standard deviation of the noise, we set s = 1, and rewrite the equation in terms of the minimum deflection angle, $\beta_L$,

$$\beta_L = \frac{\pi\alpha}{4\sqrt{n}} \quad \text{(A3)}.$$

Note that Equation A3 is the minimum deflection angle acquired via a normalized difference map of a split detector when illuminated by a circular beam illumination. In our approach, we use a modified version for a square beam illumination that matches with our simulations in Appendix B – the difference between the two approaches is a 20% correction factor, but the other scalings remain unchanged and the numerical simulation is greatly accelerated.

We also know the beam deflection angle is given as:

$$\beta_L = \frac{e\lambda B_0 t}{h} \quad \text{(A4)},$$

where $B_0$ is the local magnetic field, $\lambda$ is the wavelength of the electron beam, t is the sample thickness and h is Planck's constant. We can therefore know the minimum detectable magnetic field $B_0$:

$$B_0 = \frac{\pi h\alpha}{4e\lambda t\sqrt{n}} \quad \text{(A5)}.$$

To be able to have a direct field sensitivity comparison with other magnetic imaging techniques such as superconducting quantum interference device (SQUID) and nitrogen vacancy (NV) centers[61], we have to normalize the time interval/bandwidth for taking measurements on LSTEM so that neither our sensitivity nor the corresponding dose is dependent of dwell time. Normalizing out the time interval we get the field sensitivity $B_t$ in terms $\frac{T}{\sqrt{Hz}}$ , as:

$$B_t = B_0\sqrt{\tau} = \frac{\pi h\alpha}{4\sqrt{e}\lambda t\sqrt{I}} \quad \text{(A6).}$$

We can see from (A6) that the field sensitivity scales inversely with square root of electron beam current and scales inversely proportional with sample thickness for thin samples. The field sensitivity for our square beam has a numerical prefactor of 1 vs $\frac{\pi}{4}$ for the round beam, a 20% difference (Appendix B) .

**Appendix B**: Determining Magnetic Field Detection Sensitivity: Numerical Model

In our numerical simulation, we assume a one-dimensional detector, represented as the vector form of $\vec{v}$. Our detector, $\vec{v}$, is then entirely illuminated by an electron beam represented by a top hat vector, $\vec{g}$ which has the same size as our detector vector. For a detector with j number of pixels, $\vec{v}$ is composed of j number of indexes that starts from $(-\frac{j-1}{2})$ and have an increasing spacing of 1 so that the indexes are centered about the origin. For example, for a two-pixel detector, $\vec{v} = [-\frac{1}{2}, \frac{1}{2}]$. The top hat vector has a uniform distribution with the sum of components to be 1, thus for a j-pixel detector every component of $\vec{g}$ is equal to $\frac{1}{j}$. To convert our top hat distribution into the number of electrons distributed on the detector, we multiply the top hat vector, $\vec{g}$ with the electron dose, $n$ , as: $n * \vec{g} = [\frac{n}{j}, \frac{n}{j}, \ldots, \frac{n}{j}]$. However, real electron probes have intensity variations, thus to reflect a real electron probe, we added Poisson noise on top of our top hat distribution. To get statistics from the Poisson noise, we generate Poisson noise on N number of samples which have the same electron dose distribution on the same number of pixel detector, and a sample of the electron probe with noise is represented as $\vec{p}$.

Using our model, we calculate the COM, for the k-th sample that has an electron beam illumination $\overrightarrow{p_k}$, as $\mu_k$:

$$\mu_k = \frac{\overrightarrow{p_k} \cdot \vec{v}}{sum(\overrightarrow{p_k})} \quad \text{(B1).}$$

Here, $\mu_k$ represents the COM on our detector, $\vec{v}$, in units of pixels. We find the mean, $\underline{\mu}$, from all N samples as:

$$\underline{\mu} = \frac{\sum_{k=1}^{\#\ of\ samples} \mu_k}{N} \quad \text{(B2).}$$

Using Equation B1 and B2, we calculate the standard deviation, $\sigma$, as

$$\sigma = \sqrt{\frac{1}{N-1} \sum_{k=1}^{N} \left(\mu_k - \underline{\mu}\right)^2} \quad \text{(B3).}$$

$\sigma$ is the noise level of COM measurements, in the unit of pixels. To convert the noise level to units of deflection angle, we need to calibrate the pixel with our semi-convergence angle $\alpha$. The calibration factor is different for different numbers of pixels on our detector, since for the same convergence angle, as pixel number increases, the size of each pixel relative to the size of the electron beam is reduced. To counteract this effect, we must calibrate $\sigma$ so that our noise measurements with different numbers of pixels on our detector are consistent with the unit of angles (radians). To figure out what the calibration factor is for a j-pixel detector, we calculate the COM for a signal of $\beta_L$ deflection angle. The example for a 2-pixel detector case is shown in Fig. B1.

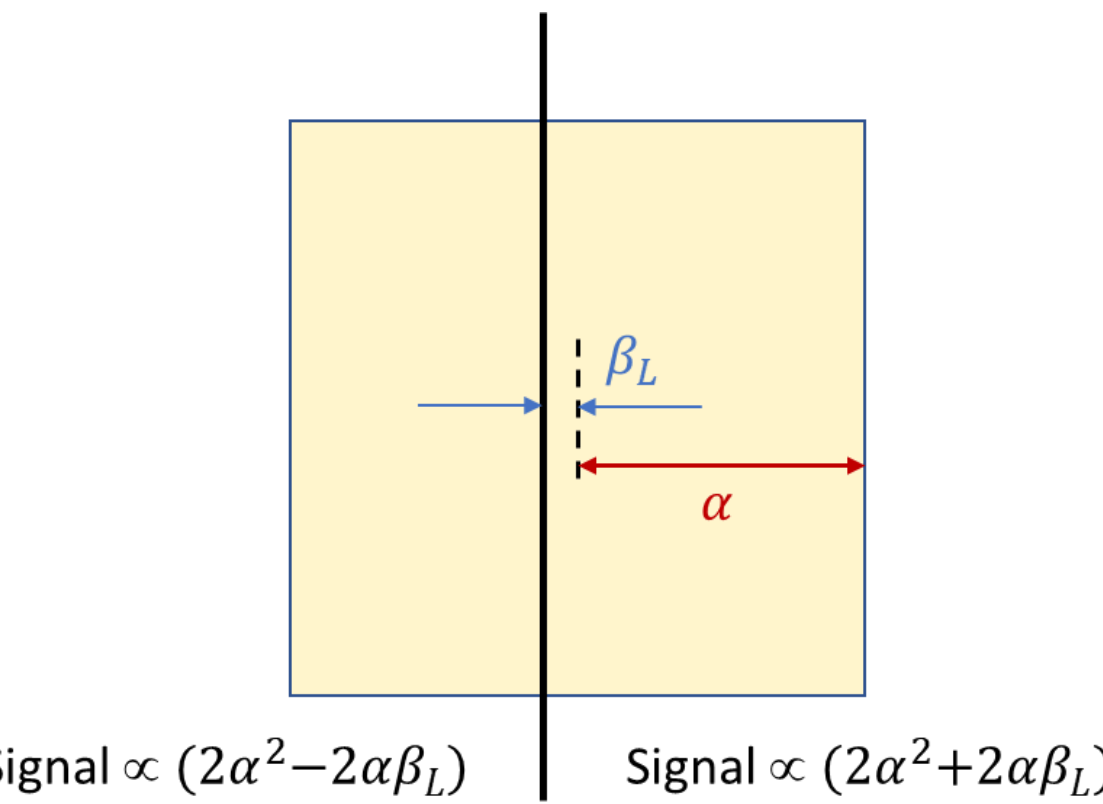

FIG. B1. The shift of the probe intensity distribution on a two-pixel detector assuming a square illumination for a semi-convergence angle $\alpha$ and a small Lorentz deflection angle $\beta_L$. The associated minimum deflection angle with electron dose n is $\beta_L = \frac{\alpha}{\sqrt{n}}$.

When there is a small $\beta_L$ deflection compared to $\alpha$ on a j-pixel detector, the signal of probe shift will only cause intensity change on the first and the last detector pixels. We know the first and last pixel indexes for a j-pixel detector are $(-\frac{j-1}{2})$ and $(\frac{j-1}{2})$, so the COM for a j-pixel detector in unit of pixel is:

$$\beta_{COM} = \frac{(j-1)\beta_L}{2\alpha} \quad \text{(B4).}$$

Therefore, the calibration factor in the unit of (rad/pixel) for a j-pixel detector is:

$$F = \frac{\beta_L}{\beta_{COM}} = \frac{2\alpha}{(j-1)} \quad \text{(B5).}$$

We perform the calibration for $\sigma$ and convert it to the minimum deflection as:

$$\beta_L = \sigma \cdot \frac{2\alpha}{(j-1)} \quad \text{(B6),}$$

where $j$ is the number of detector pixels and $\alpha$ is the semi-convergence angle of our electron beam.